\documentclass[aps,prl,twocolumn,floats,amsmath,amssymb,showpacs,showkeys,preprintnumbers,floatfix]%
{revtex4-1}

\usepackage{graphicx}
\usepackage{dcolumn}
\usepackage{color}

\definecolor{GreenOlive}{rgb}{0.1,0.5,0.1}

\begin{document}

\title{Stochastic Cascade Amplification of Fluctuations}

\author{ Michael Wilkinson$^1$, Robin Guichardaz$^2$,  Marc Pradas$^1$, and Alain Pumir$^{2,3}$}
\affiliation{$^1$ Department of Mathematics and Statistics,
The Open University, Walton Hall, Milton Keynes, MK7 6AA, England,\\
$^2$ Laboratoire de Physique,
 Ecole Normale Sup\'erieure de Lyon, CNRS, Universit\'e de Lyon,
 F-69007, Lyon, France, \\
$^3$ Max-Planck Institute for Dynamics and Self-Organisation,
D-37077, G\"ottingen, Germany 
}

\begin{abstract}

We consider a dynamical system which has a stable attractor
and which is perturbed by an additive noise. Under some quite 
typical conditions, the fluctuations from the attractor are intermittent 
and have a probability distribution with power-law tails. We show 
that this results from a stochastic cascade of amplification of 
fluctuations due to
transient periods of instability. The exponent of the power-law is interpreted
as a negative fractal dimension.

\end{abstract}

\pacs{05.10.Gg,05.40.-a,05.45.Df}

\maketitle

Intermittency refers to the alternation of long periods of rest, followed
by bursts of strong activity. This phenomenon plays a crucial role in 
the dynamics of a wide range of physical systems, for example in fluid 
dynamics~\cite{Reynolds83,ShrSig00}, astrophysics~\cite{Priest02}, 
condensed matter~\cite{FKJM08}, as well as in physiology~\cite{Krahe04}. 
Theoretical explanations proposed in the case of deterministic dynamical 
systems with a few degrees of freedom 
assume that the system is close to the 
transition to chaotic behaviour~\cite{Pom+80,Ott+94,Ven+96,Ott02} . 
The description of intermittency in extended
dynamical systems remains a challenging issue~\cite{Pom86,CM88}. \par
The importance of noise in the dynamics leading to intermittency has been 
established in pipe flows~\cite{Eckhardt07} and spatially extended systems which
are close to the instability onset~\cite{Pra+11,Pra+12}, and is likely to play an important
role in other physical systems as well. This article is concerned 
with the influence of noise on intermittency in simple model systems.

Here, we consider a universal form of intermittency arising 
from the addition of noise to a dynamical system which has 
stable dynamics. It arises for dynamical systems which are
(i) {\it non-autonomous}, and (ii) {\it stable}, in the sense that in 
the absence of noise, 
solutions converge to a point attractor. It is the interaction
between the noise and the fluctuating environment 
which generates intermittency, via a mechanism of stochastic
amplification.
In one spatial dimension we consider equations of motion in the form:
\begin{equation}
\label{eq: 1.1}
\dot x=v(x,t)+\sqrt{2D} \eta(t),
\end{equation}
where $v(x,t)$ is a velocity field fluctuating in space and time, 
$\eta(t)$ is a white noise signal with 
$\langle \eta(t) \rangle=0$ and 
$\langle \eta(t)\eta(t') \rangle=\delta (t-t')$, 
and $D$ is the diffusion coefficient of the corresponding Brownian 
motion, which we assume to be small 
($\langle X\rangle$ denotes the expectation value of $X$).
In the deterministic case ($D=0$) the system is characterised by 
its Lyapunov exponent $\lambda$. If $\lambda<0$, the system is
stable, so all the trajectories of the system converge to an attractor \cite{Wil+03}
when $D=0$. 
If the system is unstable, $\lambda>0$, there is typically a strange
attractor \cite{Ott02}.

In the presence of noise ($D \ne 0$), however, the 
trajectories do not remain on the attractor of a stable system. As a result, the
separation of two trajectories, denoted as $\Delta x$, can have 
large excursions away from $0$, leading to tails of the
probability distribution function (PDF) $P_{\Delta x}$ 
(Throughout we denote by $P_X$ the PDF 
of the quantity $X$).

The PDF $P_{\Delta x}$ may be expected to be well approximated by a 
Gaussian distribution. In the case of a stable autonomous system
the motion in the vicinity of the attractor satisfies 
$\dot{x} = \lambda x+\sqrt{2D}\eta (t)$ with $\lambda<0$,
which is an Ornstein-Uhlenbeck (OU) process~\cite{Uhl+30}.
In this case the deviations from the attractors do have a Gaussian
distribution, with a variance $D/|\lambda|$, so the deviations $\Delta x$
between two trajectories also have a Gaussian PDF, with a variance 
$2 D/|\lambda|$.

\begin{figure}[t]
\includegraphics[width=0.40\textwidth]{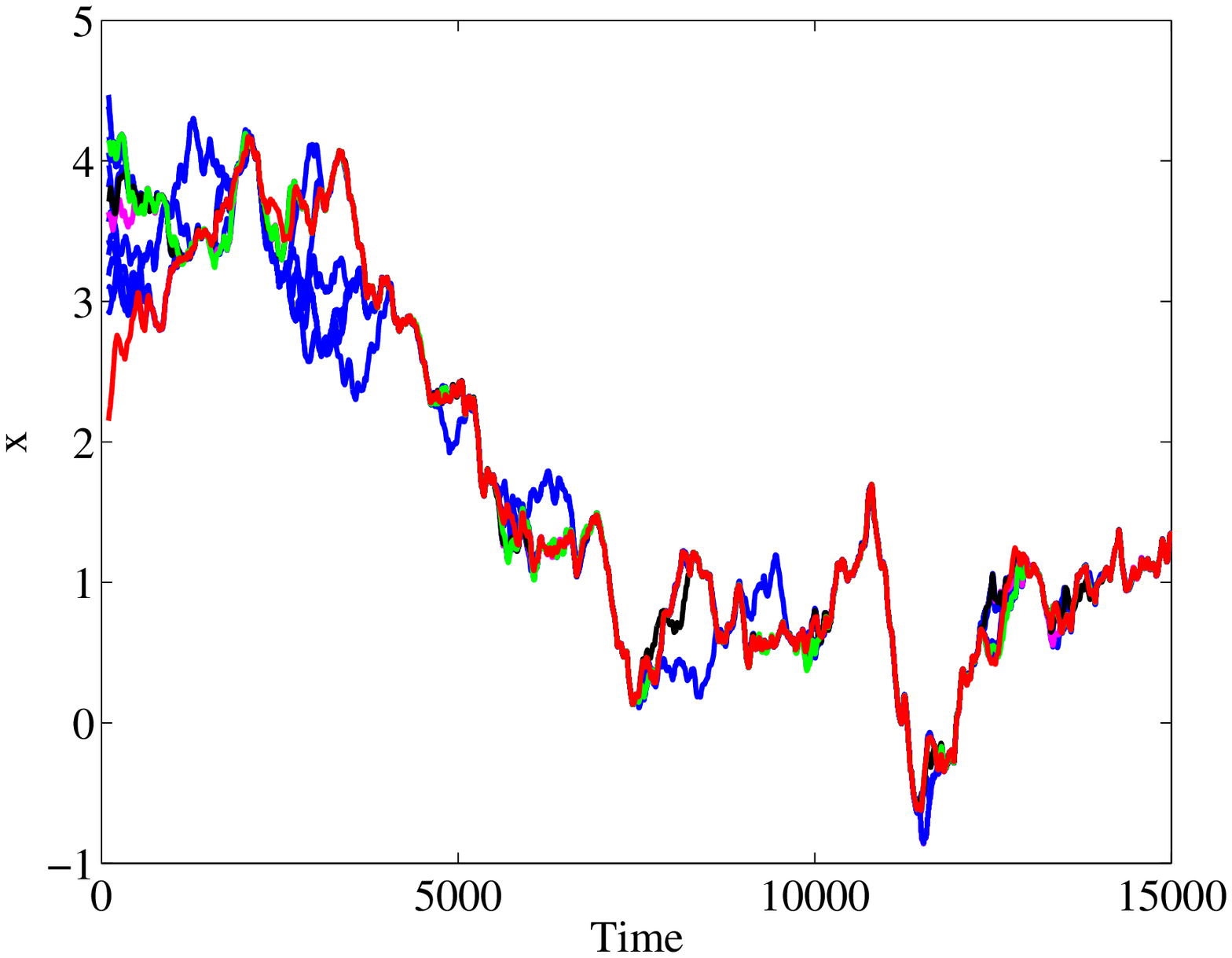}
\includegraphics[width=0.40\textwidth]{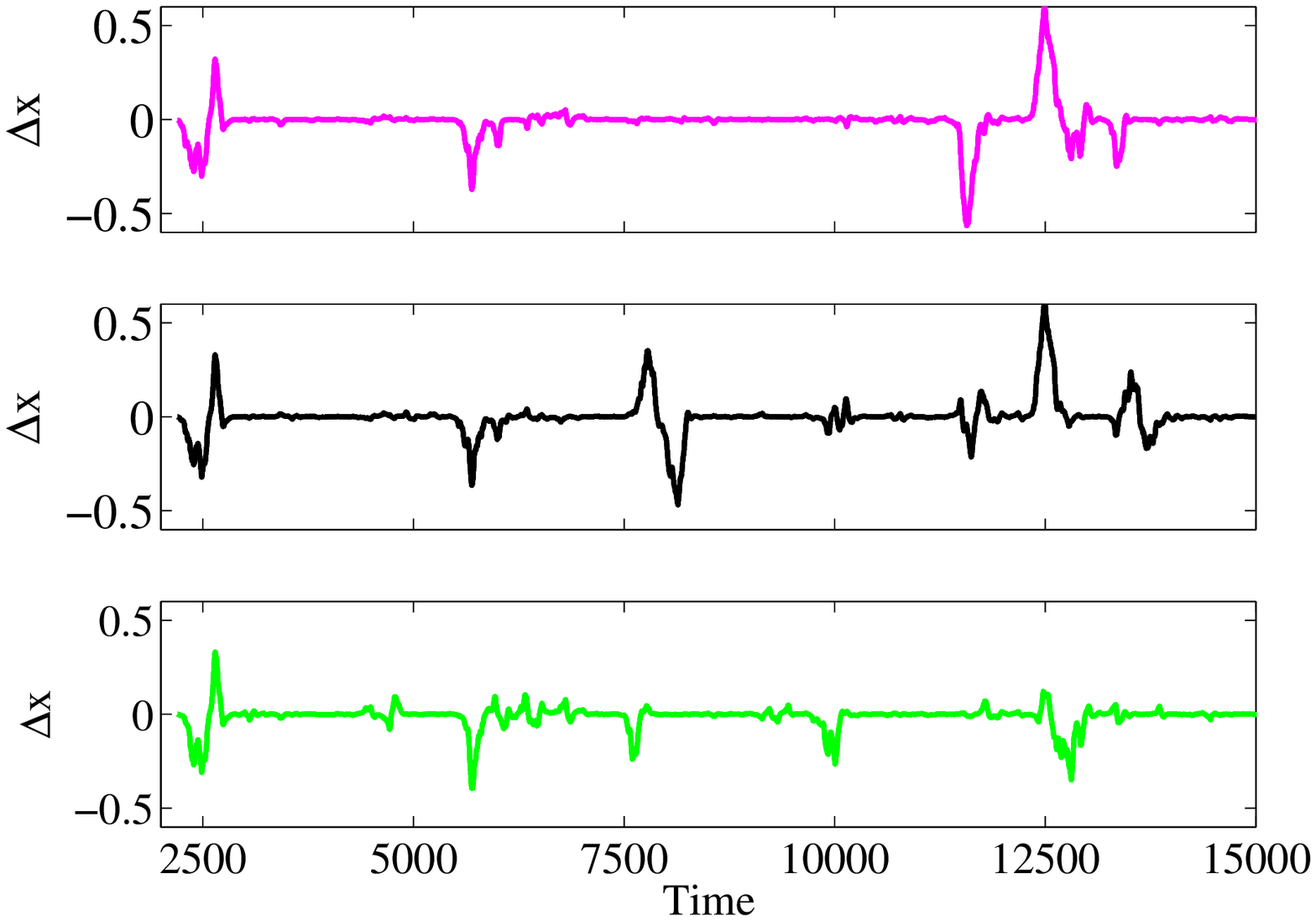}
\caption{
(Color online). 
{\bf a} Set of trajectories of 
a model of colloidal particles in suspension, Eqs.~(\ref{eq: 3.1}).
Different trajectories separate and recombine. 
{\bf b} Intermittent separation of pairs of trajectories $\Delta x(t)$.
The parameters are $\gamma= 0.05$, $\ell_\text{c}=0.08$, $A=0.02$, $D=10^{-8}$.
\label{fig: 1}
}
\end{figure}

\begin{figure}[t]
\includegraphics[width=0.40\textwidth]{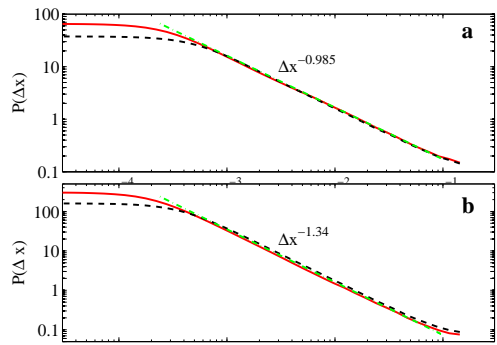}
\caption{(Color online). The probability distribution $P_{\Delta x}$ for the 
model (\ref{eq: 3.1}), for two values of $D$ ($D = 3\times10^{-10}$, full curve,
and $D=10^{-9}$, dashed curves, showing fits by a power law $|\Delta x|^{-(\alpha+1)}$.
The parameters are {\bf a} $\gamma= 0.3125$, $\ell_\text{c}=0.08$, $A=0.02$
{\bf b} $\gamma=0.375$, $\ell_\text{c}=0.08$, $A=0.02$.
\label{fig: 2}
}
\end{figure}

We demonstrate here the existence of a generic class of models 
for which the distribution $P_{\Delta x}$ is \emph{non-Gaussian} and 
has power-law tails over a range of values of $\Delta x$:
\begin{equation}
\label{eq: 1.5}
P_{\Delta x}\sim |\Delta x|^{-(1+\alpha)},
\end{equation}
when $\Delta x$ is large compared to $\sqrt{D/|\lambda|}$. 
Note that when $\alpha > 0$, the asymptotic form (\ref{eq: 1.5}) 
{\it cannot} extend all the way to $\Delta x = 0$.
The large excursions of $\Delta x(t)$ described by the power-law tails
are a manifestation of the phenomenon of intermittency, whereby close-by
trajectories are separated by the combined effect of the noise and fluctuating
environment.
This is illustrated in Fig.~\ref{fig: 1}, which shows 
individual trajectories and their differences for
a simple model  
discussed in detail below (see Eqs.~(\ref{eq: 3.1}) and (\ref{eq: 3.2})).
Figure \ref{fig: 2} shows the corresponding power-law distribution of $\Delta x$.

In this paper we explain the origin of this intermittency
and show how it can be quantified by analysing an equation which determines the exponent
$\alpha$. We show that there are universal aspects to these fluctuations which 
arise because the mechanism for producing the large fluctuations is independent 
of the mechanism which seeds them. The power-law (\ref{eq: 1.5}) 
is an emergent property, in the sense that it arises for generic equations 
of motion which do not themselves contain non-integer exponents. 
We also observe that statistics of the temporal variation of the intermittent signal, 
such as the waiting times distribution, follows a power-law behaviour. 

Beyond the point at which the underlying system
becomes unstable, the system has a
strange attractor, where phase points cluster on a fractal measure~\cite{Ott02}. 
As $\lambda$ approaches zero from below, the exponent $\alpha$ 
in Eq.~(\ref{eq: 1.5}) approaches zero from above. When $\lambda>0$, the two-point correlation 
function of the strange attractor, $g(\Delta x)$, has a power-law dependence:
\begin{equation}
\label{eq: 1.6}
g(\Delta x)\sim |\Delta x|^{D_2-1},
\end{equation}
where $D_2$ is the correlation dimension~\cite{Gra+84}. 
Equations (\ref{eq: 1.5}) and (\ref{eq: 1.6}) have 
the same structure with the exponents related by 
 \begin{equation}
 \label{eq: 1.7}
 \alpha=-D_2
 \end{equation}
so that normalisable distributions of fluctuations correspond to negative 
values of $D_2$. Equation (\ref{eq: 1.5}) therefore gives a physical 
meaning to a negative fractal dimension, but we should emphasise the 
difference between the interpretation of the two exponents. Equation 
(\ref{eq: 1.6}) describes the pair correlation function for a system 
with no added noise, whereas Eq.~(\ref{eq: 1.5}) describes the 
tail of the PDF for separations when a small noise signal
is added. The formal relation between our exponent $\alpha$ and the 
correlation dimension $D_2$ allows us to use recent advances 
in computing the correlation dimension~\cite{Wil+10,Gus+11,Wil+12} 
to quantify our exponent $\alpha$. An 
alternative definition of negative fractal dimension 
has been offered in \cite{Man90}.

We now explain the power-law tails by introducing a \emph{cascade amplification mechanism}.  
Consider the linearisation of Eq.~(\ref{eq: 1.1}) to give the separation between
two nearby trajectories:
\begin{equation}
\label{eq: 2.1}
\delta \dot x=Z(t) \delta x+2\sqrt{D} \eta(t),
\end{equation}
where 
\begin{equation}
\label{eq: 2.2}
Z(t)=\frac{\partial v}{\partial x}(x(t),t).
\end{equation}
Note that when $D=0$, $Z(t)$ is the logarithmic derivative of the 
separation $\delta x(t)$, and that its expectation value is the Lyapunov exponent:
\begin{equation}
\label{eq: 2.3}
Z(t)=\frac{\delta \dot x}{\delta x}, \qquad  \lambda=\lim_{t\to \infty}\frac{1}{t}\int_0^t{\rm d}t'\ Z(t')
= \langle Z(t)\rangle
\ .
\end{equation}
We can think of $Z(t)$ as being an \emph{instantaneous Lyapunov exponent}.  
In the case of autonomous systems with an attractor, the attractor must be a 
fixed point in phase space, and $Z(t)$ approaches a constant $\lambda<0$ 
as $t\to \infty$. In this case the fluctuations are described by an OU 
process and the distribution $P_{\Delta x}$ is Gaussian.  
In cases where the dynamical system is non-autonomous, 
$Z(t)$ need not approach a constant value. If the external driving 
is a stationary stochastic process, $Z(t)$ is a fluctuating quantity with stationary
statistics. The origin of the power-law tails 
described by (\ref{eq: 1.5}) is that the fluctuations are amplified during 
periods when $Z(t)>0$. 
This noise amplification is independent of the initial amplitude, because
the fluctuating quantity $Z(t)$ acts multiplicatively in Eq.~(\ref{eq: 2.1}).
This leads to a stochastic cascade amplification process, whereby large 
amplitude fluctuations are built up by a succession of periods where 
$Z(t) > 0$.
The power-law tail in the fluctuation distribution 
arises whenever $ Z(t)$ 
is positive for some intervals of time, however short. 

In order to quantify this picture, let us consider the dynamics of the 
fluctuations in a logarithmic variable
\begin{equation}
\label{eq: 2.4}
Y={\rm ln}(\Delta x)
\ .
\end{equation}
Consider the tail of $P_{\Delta x}$, where the fluctuations
are much larger than the driving noise, so that the term $\sqrt{2D} \eta(t)$ 
in (\ref{eq: 2.1}) can be neglected. In this limit the equation of motion 
for $Y(t)$ is simply $\dot Y=Z(t)$. Because the fluctuations of $Z(t)$ are
independent of $Y$, the PDF of $Y$ is expected to approach 
a limit which has translational invariance, up to a normalisation factor. This implies that:
\begin{equation}
\label{eq: 2.5}
P_Y\sim\exp(-\alpha Y)
\end{equation}
where the constant $\alpha$ depends on the 
statistics of $Z(t)$ \cite{Wil+12}. The corresponding 
PDF of $\Delta x$ [Eq.~(\ref{eq: 1.5})] is then obtained 
by writing the change of variable  ${\rm d}P=P_Y{\rm d}Y=P_{\Delta x}{\rm d}\Delta x$.
Note that when $Y \to -\infty$, or alternatively, $\Delta x \to 0$, the noise term
dominates in Eq.~(\ref{eq: 2.1}), so Eq.~(\ref{eq: 1.5}) does not apply. This
prevents any difficulty with the divergence of $P_Y$.
In fact, the role of the noise term reduces to providing 
a natural cutoff at small separations ($\Delta x \to 0$), 
thus preventing potential normalisation problems. 
It also follows that the exponent $\alpha $ is independent of $D$, provided that $D>0$.

Next we describe a concrete and physically 
important example of a system that produces
fluctuations described by (\ref{eq: 1.5}). 
This is provided by \emph{colloidal particles in a turbulent fluid flow}, 
with velocity field $\mbox{\boldmath$u$}(\mbox{\boldmath$x$}(t),t)$. The 
motion of small particles suspended in the flow is determined by viscous
drag, which makes their velocity $\mbox{\boldmath$v$}(t)$ relax to that 
of the surrounding fluid.
When the particles have a density which is much higher than that
of the fluid in which they are dispersed, the viscous drag is 
proportional to the difference between the particle velocity 
and the fluid velocity at its current location
(limitations of the model
are discussed in \cite{Gat83,Max+83}). 
The problem studied 
in this paper is based upon a one-dimensional version of this model,
which includes Brownian motion of the particles:
\begin{subequations}\label{eq: 3.1}
\begin{eqnarray}
\dot{x} &=&v+\sqrt{2D} \eta (t), \label{eq: 3.1a} \\
\dot{v} &=&\gamma[u(x,t)-v], \label{eq: 3.1b}
\end{eqnarray}
\end{subequations}
where $\gamma$ is proportional to the viscosity of the fluid.
We consider a random velocity field with a vanishingly 
small correlation time, with statistics given by
\begin{subequations}\label{eq: 3.2} 
\begin{eqnarray}
\langle u(x,t)\rangle &=& 0, \label{eq: 3.2a} \\
\langle u(x,t)u(x',t')\rangle &=& A^2\exp\left[-\frac{(x-x')^2}{\ell_\text{c}^2}\right]\delta (t-t').\qquad
\label{eq: 3.2b} 
\end{eqnarray}
\end{subequations}
All of the explicit work shown in this paper is based on 
Eqs.~(\ref{eq: 3.1}) and (\ref{eq: 3.2}).
This model has two very different random elements.
The random velocity field $u(x,t)$ is the same for all trajectories, whereas 
the Brownian noise $\eta(t)$ has a different realisation for each particle 
trajectory. 
This model has been extensively analysed 
for $D=0$. The Lyapunov exponent, obtained in \cite{Wil+03}, 
was found to be negative for sufficiently 
large values of $\gamma$, with the attractor not being  a 
fixed point, but a random walk. The separation 
of trajectories was analysed by \cite{Der+07} and the correlation 
dimension was investigated in \cite{Gus+11} for the 
case where $D_2\ge 0$. When $D\ne 0$, we find that the 
deviations of the trajectories are found 
to exhibit a power-law tail in their PDF, described by Eq.~(\ref{eq: 1.5}), see  
Fig.~\ref{fig: 2}.

Consider the linearisation of Eqs.~(\ref{eq: 3.1}):
\begin{eqnarray}
\label{eq: 3.3}
\delta \dot x &=&\delta v +2\sqrt{D}\eta(t),
\nonumber \\
\delta \dot v &=&\gamma\left[S(t)\delta x-\delta v\right],
\end{eqnarray}
where $S(t)=\frac{\partial u}{\partial x}(x(t),t)$ is the velocity gradient at the 
position of a particle, which can be modelled by a white noise signal with 
diffusion coefficient ${\cal D}$:
\begin{equation}
\label{eq: 3.5}
S(t)=\sqrt{2{\cal D}}\zeta(t)
\ ,\ \ \ \ 
{\cal D}=\frac{A^2}{\ell_\text{c}^2}
\end{equation}
where $\zeta(t)$ is independent of $\eta(t)$ but has the same
statistical properties.
 
When $D=0$, from Eqs.~(\ref{eq: 3.3}) and (\ref{eq: 3.5}) we obtain the following 
equation of motion for $Z(t)$ (previously obtained in \cite{Wil+03}):
\begin{equation}
\label{eq: 3.6}
\dot Z=-\gamma Z-Z^2+\sqrt{2{\cal D}}\zeta (t).
\end{equation}
We wish to determine the PDF for $Y$ in the form (\ref{eq: 2.5}).
From Eq.~(\ref{eq: 3.6}) we can construct a Fokker-Planck equation
for the joint probability density $\rho(Y,Z)$ of $Y$ and $Z$, and seek a solution in the 
form $ \rho(Y,Z)=\exp(-\alpha Y)\rho_Z(Z)$ for consistency with (\ref{eq: 2.5}). 
Following an approach described in \cite{Wil+10}, this leads 
to a differential equation for $\alpha$ in the form 
 \begin{equation}
 \label{eq: 3.7}
 \hat F \rho_Z(Z)+\alpha Z\rho_Z(Z)=0
 \end{equation}
where $\hat F$ is a differential operator defined by writing 
 \begin{equation}
 \label{eq: 3.8}
 \hat F\rho(Z)=\frac{\partial }{\partial Z}\left[(\gamma Z+Z^2)
 +{\cal D}\frac{\partial}{\partial Z}\right]\rho(Z)
 \ .
 \end{equation}
 Because $\partial/\partial Z$ is a left-factor of $\hat F$, 
any normalisable solution of (\ref{eq: 3.7}) with $\alpha\neq 0$ must satisfy:
 \begin{equation}
 \label{eq: 3.9}
 \int_{-\infty}^\infty{\rm d}Z\ Z\,\rho_Z(Z)=0
 \ .
 \end{equation}
Equation  (\ref{eq: 3.9})  must be imposed 
on any approximate solutions of (\ref{eq: 3.7}) constructed by perturbation 
theory. We remark that the integral in (\ref{eq: 3.9}) is
distinct from the Lyapunov exponent, because $\rho_Z(Z)$ 
is a distribution of $Z$ which is conditional upon the value of $Y$.

\begin{figure}[t]
\includegraphics[width=0.40\textwidth]{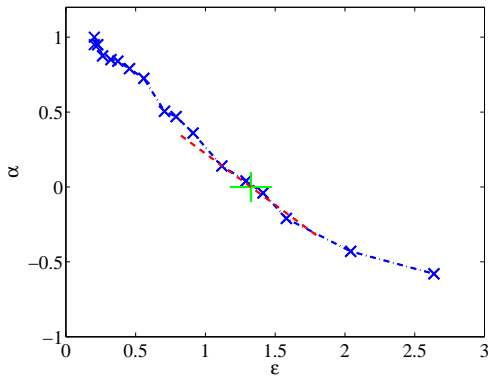}
\caption{(Color online) 
Exponent $\alpha$ defined by (\ref{eq: 1.5}) as a function 
of the dimensionless parameter $\varepsilon$ defined by (\ref{eq: 4.1}).
Note that $\alpha\to1$ as $\varepsilon\to 0$ and that in the vicinity of the 
critical point $\varepsilon_{\rm c}$ where $\lambda=0$, we have 
$\alpha \sim K(\varepsilon_{\rm c}-\varepsilon)$. The dashed line uses 
our theoretical value for the coefficient $K$ and the cross 
shows the critical point location.}
\label{fig: 3}
\end{figure}

In order to facilitate the analysis of (\ref{eq: 3.7}), we 
replace $Z$ by a scaled
variable $x$, and introduce a dimensionless parameter $\varepsilon$: 
\begin{equation}
\label{eq: 4.1}
x(t)=\sqrt{\frac{\gamma}{{\cal D}}}Z(t)
\ , \ \ \ \ 
\varepsilon=\sqrt{\frac{{\cal D}}{\gamma^3}}
\ .
\end{equation}
With these definitions, Eq.~(\ref{eq: 3.8}) is replaced by
\begin{equation}
\label{eq: 4.2}
\frac{\partial}{\partial x}\left[x+\varepsilon x^2+\frac{\partial}{\partial x}\right]\rho(x)+\alpha  \varepsilon   x  \rho(x)=0
\ .
\end{equation}
Our numerical results indicate that the exponent $\alpha$, determined
by directly computing $P_{\Delta x}$, is a function of 
the scaling variable $\varepsilon$, as illustrated in Fig.~\ref{fig: 3}.
We consider two different perturbative approaches to determining 
$\alpha $ as a function of $\varepsilon$. The first is to make an expansion
about $\varepsilon =0$. This can be done by following the method 
discussed in \cite{Wil+10}. The series 
expansion of $\alpha$ has only one non-zero term: $\alpha=1$, and 
all of the coefficients of higher powers of $\varepsilon$ are identically 
zero \cite{Meh}. The implication is that $\alpha$ has a non-analytic dependence
upon $\varepsilon$. 

An alternative approach is to make a perturbative expansion 
about the critical point where the Lyapunov exponent changes 
sign. For the model underlying (\ref{eq: 3.1}), this occurs at 
$\varepsilon_{\rm c} \approx 1.33\ldots$. We follow an approach used 
in \cite{Gus+11} (see also \cite{Sch+02}).  
To leading order in $\varepsilon-\varepsilon_{\rm c}$ we obtain
\begin{equation}
\label{eq: 4.3}
\alpha=K(\varepsilon_{\rm c}-\varepsilon)
\ . 
\end{equation}
Using the approach described in \cite{Gus+11} and \cite{Sch+02} the coefficient $K$ is 
expressed in terms of finite-dimensional integrals, whose numerical 
evaluation leads to $K \approx 0.688$. This value is found to be in good 
agreement with the results shown in Fig. \ref{fig: 3}. 

\begin{figure}[t]
\includegraphics[width=0.48\textwidth]{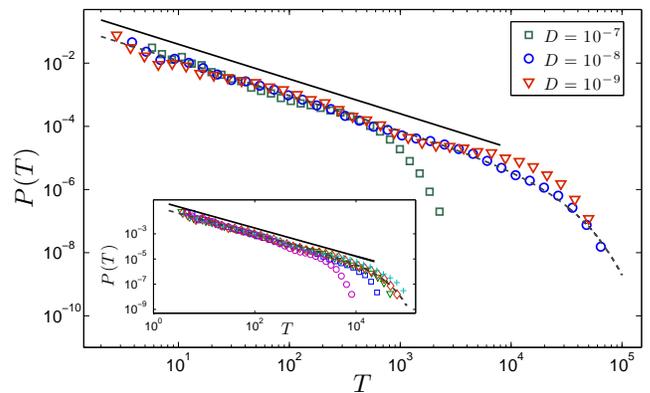}
\caption{(Color online) PDF of the waiting times $T$ for different values of the noise
 intensity $D$. The dashed line corresponds to a data fit to the function 
 $P_T = N T^ {-\tau}\exp{(-b T)}$ with $\tau = 1.12\pm 0.07$, and 
 the solid line corresponds to a power law with exponent $-1.12$. The inlet 
 shows results obtained by choosing different values of the threshold $c_\text{th}$ 
($0.002 \le c_\text{th} \le 0.02$)
for the case of $D= 10^{-8}$. 
The parameters of the simulation are $\gamma = 0.05$, $A = 0.02$ and $\ell_\text{c} = 0.08$.
\label{fig: 4}
}
\end{figure}

We have shown that intermittency in our model leads to 
power-law behaviour of $P_{\Delta x}$. 
We now investigate the temporal variation 
of the signal $\Delta x(t)$. The intermittency of $\Delta x(t)$ shown in Fig.~\ref{fig: 1} 
can be characterised by considering the distribution of waiting time intervals $T$ over 
which $\Delta x(t)$ remains below a defined threshold, say $c_\text{th}$. Figure \ref{fig: 4} 
shows the PDF $P_T$ of the waiting times $T$ for three different cases of noise intensity, namely 
$D = 10^{-7}$, $10^{-8}$, and $10^{-9}$, and for the case 
where $\varepsilon \approx 1.118$. 
We observe that $P_T$ follows a power-law 
behaviour with an exponential decay at long times which can be fit to 
the function  $P_T = N T^ {-\tau}\exp{(-b T)}$ 
with an exponent $\tau =1.12\pm 0.07$, independent of $D$.  
Note that varying $D$ only affects the cut-off value of the 
exponential tail, i.e.~the value of the parameter $b$: increasing $D$
decreases the range of values of $T$ over which $P_T$ has a power-law 
dependence. The inset of Fig.~\ref{fig: 4} shows that the choice of the threshold 
$c_\text{th}$ does not affect the exponent of the power-law regime, but only modifies
$b$. It is important to remark that none of the mechanisms explaining 
the power-law distribution of inter-burst times observed in other physical 
systems (see e.g.~\cite{Vaz+06,Lau+09,Pra+11})  is applicable in the present problem.

To conclude, we have explained and characterised a class of intermittent 
fluctuations in dynamical systems. They differ from the usual
types of intermittency considered in low-dimensional systems, 
in that they arise when the equations of motion have additive 
noise, but the underlying dynamical system is stable 
(i.e. has a negative Lyapunov exponent). We have shown
using symmetry arguments, and the assumption that the instantaneous
Lyapunov exponent has positive fluctuations, that
the intermittency is characterised by a power-law distribution 
of the magnitude of the fluctuations, and we have discussed 
perturbative methods for estimating the exponent $\alpha$. 
We have also presented evidence that there are power law distributions
of other quantities, such as the inter-burst times, which is a 
feature common to many different intermittent systems. The 
stochastic cascade amplification
mechanism for producing these fluctuations is generic, so that they should be 
observable in a wide class of systems.

\end{document}